\documentclass[12pt]{article}
\usepackage{graphicx}
\usepackage{color}
\newcommand{\beq}{\begin{equation}}
\newcommand{\eeq}{\end{equation}}
\newcommand{\bea}{\begin{eqnarray}}
\newcommand{\eea}{\end{eqnarray}}

\newcommand{\gsim}{\lower.7ex\hbox{$
\;\stackrel{\textstyle>}{\sim}\;$}}
\newcommand{\lsim}{\lower.7ex\hbox{$
\;\stackrel{\textstyle<}{\sim}\;$}}

\newcommand{\eod}{\end{document}}

\def\op{{\bf P}}
\def\oc{{\bf C}}
\def\ot{{\bf T}}
\def\cp{{\bf CP}}
\def\cpt{{\bf CPT}}
\def\CPV{{\bf CPV}}
\def\TV{{\bf TV}}

\usepackage{epsfig}

\definecolor{verm}{rgb}{0.8,0.1,0.0}

\begin{document}
\thispagestyle{empty}
\vspace*{-22mm}

\begin{flushright}

UND-HEP-14-BIG\hspace*{.08em}07(/UND-HEP-16-BIG\hspace*{.08em}01)\\
Version 5.0 \\

\end{flushright}

\vspace*{1.3mm}

\begin{center}
{\Large {\bf CP Asymmetries (\& T Violation) in Known Matter -- and beyond}}

\vspace*{10mm}

{\em From Roman history about data: Caelius (correspondent of Cicero) had taken a pragmatic judgment of  who was likely to win the conflict and said: 
“Pompey had the better cause, but Caesar the better army, and so I became a Caesarean.}
\footnote{`Cause' = symmetry, yet `army' = data.} 

\vspace*{10mm}

{\bf I.I.~Bigi$^a$} \\
\vspace{7mm}
$^a$ {\sl Department of Physics, University of Notre Dame du Lac}\\
{\sl Notre Dame, IN 46556, USA}

\vspace*{-.8mm}

{\sl email addresses: ibigi@nd.edu}

\vspace*{10mm}

{\bf Abstract}\vspace*{-1.5mm}\\
\end{center}
Finding \cp~violation (\CPV ) in 1964 produced a real revolution in the fundamental dynamics, although the 
community did not understand it right away. The paper by Kobayashi \& Maskawa \cite{KM} appeared 
in 1973 to describe 
\CPV~in classes of three (or more) families of quarks, non-minimal Higgs' dynamics and/or charged 
right-handed currents. The Standard Model is defined now with 
three families of quarks. It can describe the measured \cp~ \& \ot~violation in kaon and $B$ mesons at 
least as the leading source. None has been found in baryons, charm mesons, top quarks and EDMs -- except from `hot' news 
from LHCb data about {\bf T}-odd moment in $\Lambda_b^0 \to p \pi^-\pi^+\pi^-$ \cite{LHCbtalk}.  
We have failed the explain our matter vs. anti-matter huge asymmetry. Even when there is no obvious connections with that asymmetry, 
it makes sense to probe \CPV~for the signs of 
New Dynamics \& their features. Furthermore we have to measure regional \CPV~in many-body 
final states with accuracy. I discuss EDMs, axion's impact on cosmology \& its connection with Dark Matter; finally I talk about \CPV~in leptonic dynamics.

\vspace{3mm}

\hrule

\tableofcontents
\vspace{5mm}

\hrule\vspace{5mm}
\pagebreak

\begin{figure}[h!]
\begin{center}
\includegraphics[scale=0.55]{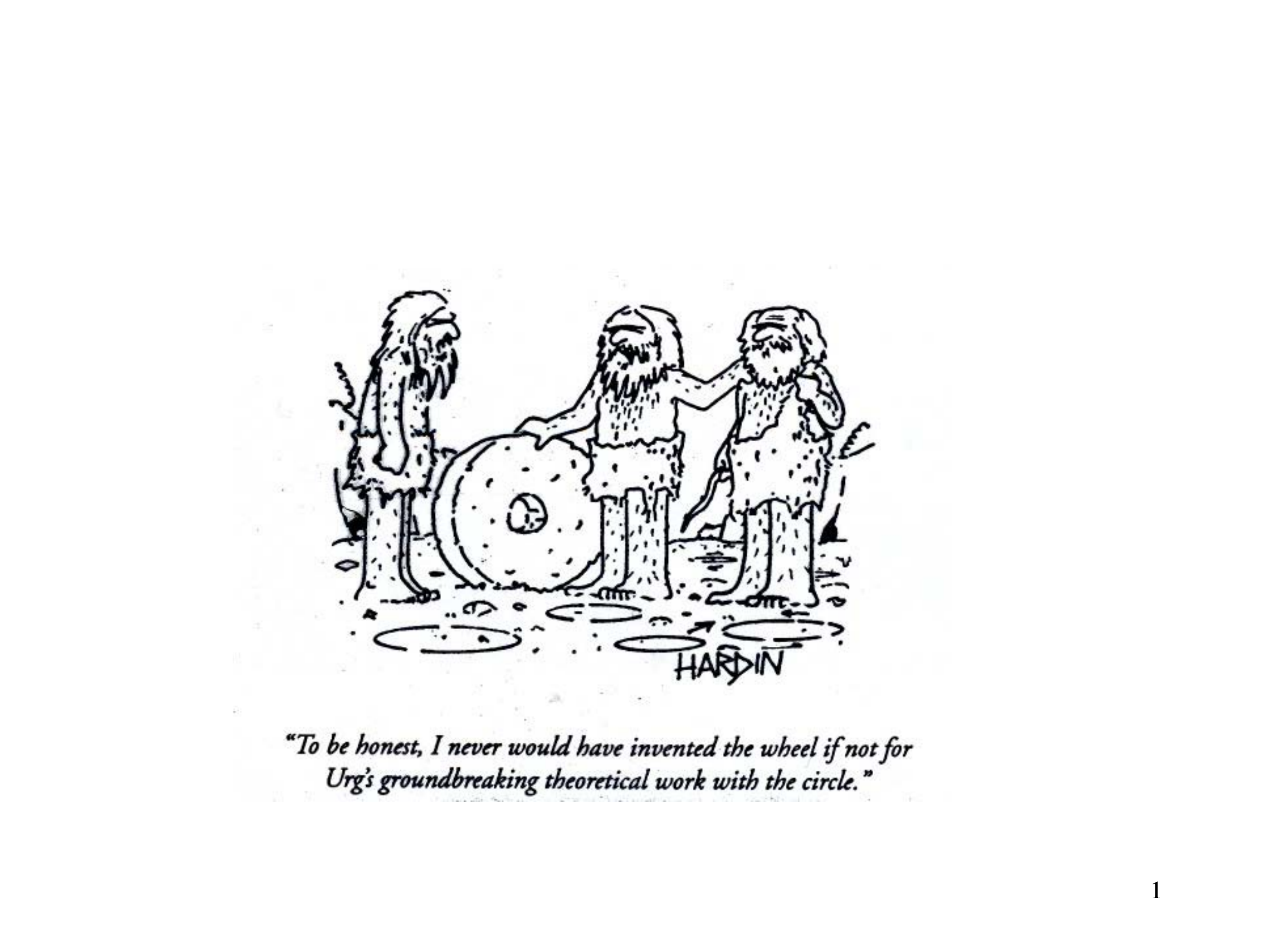}
\end{center}
\caption{" To be honest, I never would have invented the wheel if not for Urg's groundbreaking 
theoretical work with the circle." 
[A long time ago I had found this cartoon on an in-flight journal.] } 
\label{fig:WHEEL}
\end{figure}
Obviously I am a theorist working with the tools we got from quantum mechanics \& 
quantum field theories. I cannot express better -- in one sentence -- about the connection with the works of experimenters and theorists, as you see in the Fig.\ref{fig:WHEEL}  
\footnote{Not all colleagues are so polite to give such credit.}. 

\vspace{3mm}

{\large {\bf Prologue}}

\vspace{3mm}

\begin{figure}[h!]
\begin{center}
\includegraphics[width=6cm]{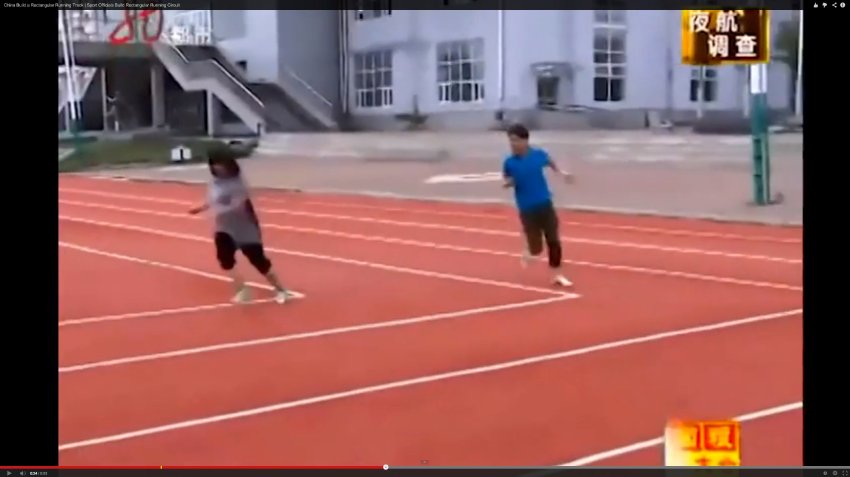} 
\end{center}
\caption{Running with discrete symmetry} 
\label{fig:CHINA}
\end{figure}

\noindent
This is a short review about \cp~violation and with some comments about the complex scenario of 
time reversal. The goal is to remember `mature' readers what they have heard before; for the `young' ones it should show the `roads' where one can learn from references in details. 
Furthermore we have to use tools based on local gauge symmetries. One can see the difference about 
local vs. discrete symmetries in real world on the Fig.\ref{fig:CHINA}, namely scenarios of physics 
vs. chemistry. 
There is a very long history on our planet. 
It has been the goal to understand fundamental dynamics: first about  `elements' and then `elementary particles'  in more \& more refined versions. Afterwards we have used the practical words 
of `high energy physics' (HEP) instead. Somewhat recently our community realized that we might barking at the wrong tree; instead we have to think about `symmetries' like local ones starting with 
QED and later about weak and strong forces. There is another class of symmetries, namely {\em discrete} ones;  
there are subclasses: 
(i) Parity (\op), charge conjugation (\oc) and time reversal (\ot); 
(ii) chiral symmetry; (iii) flavor symmetry. Those discrete symmetries are correlated due to  dynamics in important ways as discussed below in some details. 

I focus on \cp\ violation; on the other hand (broken) chiral and flavor symmetries have 
great impact on \cp\ asymmetries. I assume that the reader knows about basic tools 
for quantum mechanics \& quantum field theories including non-abelian ones, 
Einstein-Podolsky-Rosen (EPR) correlations from quantum mechanics 
\footnote{Often the word of `entanglement' has been used in this situation.}  
and the impact of symmetries like \op/\oc/\ot  
\footnote{"All animals are equal, but some animals are more equal than others!" G. Orwell, `Animal Farm'.}.  

Operators \op\ and \oc\ are unitary, namely 
${\rm \op^{\dagger}[\oc^{\dagger}]} = {\rm \op^{-1}[\oc^{-1}]}$. 
However the situation for the anti-unitary operator \ot~ is more subtle; one of the reasons  
changes initial $\rightleftharpoons$ final states: 
\beq
\langle A|{\rm \ot}^{\dagger}{\rm \ot}|B\rangle = \langle B|A\rangle
\eeq
\cpt~invariance has been assumed as usual for excellent reasons in this article.

On the other hand it can help to understand the underlying dynamics, namely Kramers'  degeneracy: 
\beq
\ot ^2 |x_1,s_1; ...;x_n,s_n\rangle = (-1)^n |x_1,s_1; ...;x_n,s_n\rangle \; ;
\eeq
i.e., ${\rm \ot}^2 = -1$ applies to a system with an odd number of same fermions and thus 
`usable'. It also helps to understand the rules of `detailed balance'.

There are more general comments about \cp\ \& \ot\ asymmetries:
\begin{itemize}
\item 
This article focuses on the weak decays of kaon, charm \& beauty hadrons and charged lepton $\tau$ with short comments about evidence of {\bf CP} 
asymmetry in $\bar \Lambda_b^0 \to p \pi^-\pi^+\pi^-$. 

\item 
I mention weak decays of top {\em quarks}, but also their production in the connection with other states at very high energy collisions like Higgs states 
\cite{TOP,CPBOOK}. 

\item 
We have to continue probing \cp\ \& \ot\ asymmetries in flavor independent transitions like in 
electric dipole moments (EDM); non-zero number has not been found yet. 

\item 
With neutrino oscillation having been established we have a subtle, but wide landscape for New Dynamics (ND) to probe. It needs long time to achieve the goal -- but there is a needed `price' 
to reach the `prize'.

\item
History shows us there was a difference between nuclear and high energy physics; however I think that is not the best road now. In particular in Europe there are groups working 
between nuclear forces and HEP called Hadrodynamics. We can see the connections between the tools produced in one section and than applied to others. It says it with different words: there are excellent reasons 
to probe fundamental forces at much higher energies -- but also to go from accuracy to precision 
at lower energies with different tools. It seems to me that we are still at the beginning of this latter road. 

\end{itemize} 

\section{History of \cp\ violation \& preview of the future}
\label{PAST}
It is possible that we live on one of many, many universes (or multi-verses). However 
our one is very special not only about the huge asymmetry of matter vs. anti-matter. 
The data tell us that baryonic (or known) matter produces around $\sim$ 4.5\%, dark matter $\sim$ 26.5 \% and vacuum (or dark) energy $\sim$ 69 \% in our universe.  
Those ratios are not very close zero or 100\% as you might guess,  
but sizable; therefore we have to deal with surprising landscape.  
Furthermore we have candidates for dark matter (like several versions of Super-Symmetry (=SUSY)). 
On the other hand we have hardly `realistic' candidates for dark energy; 
at least I am too old to spend daytime to think about vacuum energy.    

There is excellent evidence about the asymmetry of matter vs. anti-matter -- namely `our existence' on our earth.  
It was a real surprise to find that parity {\bf P} and 
charge conjugation {\bf C} are broken in charged weak forces.  
Our community 
quickly recovered that `$\tau$' and `$\theta$' -- called then -- are the same state: 
$K^{\pm}$ mesons decay to  
both parity even and odd final states. Furthermore we have neutral ones -- $\bar K^0$ 
and $K^0$ -- producing two mesons that are differentiated by their lifetimes: $K_S$ and $K_L$.  
Therefore $K_L$ was seen as parity odd mesons. Not only {\bf P} \& {\bf C} violations were found, 
but also in maximal ways, namely charged weak mesons coupling only to 
left-chiral quarks.  Also $\nu_L$ \& $\bar \nu_R$ were found , but not $\nu_R$ \& $\bar \nu_L$ for massless neutrinos. 
It is fine in a simple realization of {\bf CPT} invariance.  

A true revolution happened in 1964: it was found that $K_L$ that usually decays into three pions, 
can also -- rarely -- to two pions  
\footnote{Actually Okun stated 
in his book `Weak Interactions of Elementary Particles' published in 1963 in Russian -- 
i.e., clearly {\em before} the discovery of {\bf CP} violation (\CPV) in 1964 -- it is crucial to probe 
{\bf CP} asymmetries.}. At first 
it was suggested to introduce {\em non-linear} terms into the Schr\" odinger equation with novel  
unobserved neutral particle $U$ with $K_L \to K_S +U \to \pi \pi +U$ rather than giving up on 
{\bf CP} symmetry. More data and more thinking showed we have found \CPV~in the data.  
Then Wolfenstein gave a paper about  what is called super-weak \CPV. In my view it is not 
even a model; it is a classification for models of \CPV. 
In 1973 Kobayashi \& Maskawa gave a published paper \cite{KM}, where \CPV~can come from three classes: 
three (or more) families of quarks 
or/and charged Higgs states or/and more weak bosons with spin-one couplings to 
right-chiral bosons.   
At first some colleagues suggested the source of \CPV~comes from charged Higgs; afterwards we knew that 
we need (at least) three families of quarks for hadronic dynamics. Now we know that the CKM matrix produces 
at least the leading source of the measured {\bf CP} asymmetries in the decays of kaons and $B$ meson; that is a tested part of the SM. 

No \CPV~has been established (yet) in the dynamics of charm hadrons \& baryons in general (except our existence) and in the productions \& 
decays of top quarks (before they can produce top hadrons \cite{TOP}). 

In the SM \cp~landscape is simple for leptons: when the three neutrinos are massless, one defines 
their leptonic flavor  numbers by their couplings with charged leptons; furthermore $e$, 
$\mu$ and $\tau$ have not shown (yet) \cp~asymmetries in their decays -- except in 
$\tau^- \to \nu \bar K^0\pi^-...\to \nu K_S \pi^-...$ piggyback riding on  
$\bar K^0 - K^0$ oscillations \cite{TAUBS}.

Very short summary about past experience and prediction for the future:

(1) In the SM {\bf CP} violation is not always tiny as we found out first in neutral kaons. In the transitions of 
$B$ mesons they are large, even somewhat close to 100 \% \cite{CPBOOK}. 
The SM produces at least the leading source for those, 
but our understanding of the impact of non-perturbative QCD still is limited quantitatively. 

(2) \CPV\ was found and established in neutral kaon decays, namely indirect and direct ones in  $\epsilon_K$ and $\epsilon^{\prime}/\epsilon_K$, respectively. 
However the `job' has not been finished yet about fundamental dynamics.  When one looks 
at the triangles from the CKM matrix, one seems the impact of our understanding of $\epsilon_K$ (like in Fig.\ref{fig:CKMTRIANGLEFIT} below). 
I was told there is `soon' a chance that progress in lattice QCD will show the impact also on $\epsilon^{\prime}$.  

(3) On the other hand the CKM dynamics has nothing to do with huge asymmetries in matter vs. 
anti-matter. Therefore we still have to think \& work about this source. 

(4) Measurable {\bf CP} asymmetries need interferences of at least two amplitudes. 
Never mind that we have failed to understand matter vs. anti-matter in our universe.  
The interference can linearly depend on the amplitude of ND and thus 
allows  with much more sensitivity.

(5) Asymmetric beams of $e^+e^-$ collisions and new technologies for detectors with precision had entered a new era with the experiments 
BaBar at Stanford (U.S.A.), Belle at KEK (Japan) and now LHCb at CERN; 
it will continue with the experiment Belle-II at KEK. It is a real challenge to   
analyze huge amount of data. It is crucial to measure correlations between different final states -- 
including {\em multi}-body FS in charm \& beauty decays.

(6) On the theoretical side new tools with more accuracy to probe fundamental dynamics including operator product expansion, heavy quark expansion and lattice QCD. While the 
{\em source} of \CPV\ is 
weak forces, their {\em impact} depends on strong forces -- i.e., nonperturbative QCD. 

(7) Flavor {\em in}dependent \CPV~have been probed, in particular for EDMs in very different landscapes from 
elementary leptons to very complex states like nuclei or molecules -- and we have to continue.

(8) Based on \cpt~invariance \CPV~\&~\ot~are well connected. Of course one wants to probe 
\cpt~invariance. 
Usually it is {\em assumed} that EPR correlations  
are perfect. 

(9) There are two classes of \cp~asymmetries: 

\noindent
(i) `Indirect' \CPV~that can happen only on 
neutral mesons and need oscillations; these \cp~asymmetries depend on the time of decay; 
observables are defined by the {\em initial} FS: $K^0$ (or $K_L$), $D^0$, $B_d$ \& 
$B_s$\footnote{Often the neutral $B$ mesons are named $B^0$ \& $B^0_s$; 
I prefer to use $B=B_{u,d,s,c}$ to make it clearer.}.

\noindent
(ii) `Direct' \CPV~can be seen in the decays of hadrons (and possibly also in some them in 
production connected with other states). Its impact depends on the FS and does not depend on the time of decay. 

\noindent 
(iii) In neutral mesons one sees the interferences with both classes of \CPV. 
Their impact depends on strong final state interaction (FSI) or re-scattering based on quantum theory amplitudes. It can be described in the worlds of hadrons or quarks. 

(10) \cpt~invariance tells us that 
\CPV~is described by complex phases. However, it does not mean that all of these produce observables   
like about fermion fields and in particular about quark ones. One can change the phase of a quark field 
given CKM matrix element and rotate it away; it will re-appear in other 
matrix elements; for example: $|s\rangle \to e^{i\delta_s}|s\rangle$ leads to $V_{qs} \to e^{i\delta_s}V_{qs}$  with $q=u,c,t$. 
Kobayashi \& Maskawa showed we need three families of quarks to produce \CPV~and 
describe with a {\em single} complex phase \cite{KM} \footnote{CKM phase like the "Scarlet Pimpernel: Sometimes here, sometimes there, sometimes everywhere".}. 
In other words: one describes \CPV~in six triangles with very different patterns; however they give the same area.  
You might say that `maximal' \CPV~means 
a phase is 90$^o$. However such a statement is fallacious as said before \cite{CPBOOK}.    

\begin{figure}[h!]
\begin{center}
\includegraphics[width=8cm]{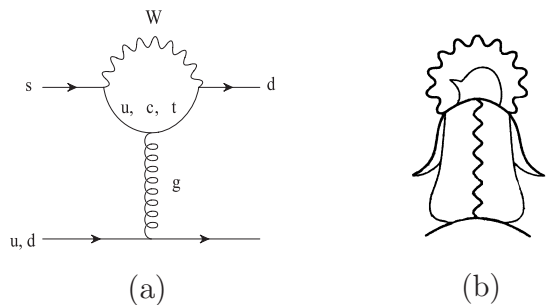} 
\end{center}
\caption{\small{Diagrams of penguin amplitudes; the picture of the (b) diagram was reproduced 
from {\em Parity} by permission of T. Muta \& T. Morozumi}} 
\label{fig:nicepeng}
\end{figure}

(11) Penguin diagrams are described in the world of quarks, gluons and weak bosons. 
Fig.\ref{fig:nicepeng}(a) sees Feynman diagrams with gluon \& $W$ gauge bosons and 
also $b$ quarks in the initial state; Fig.\ref{fig:nicepeng}(b) describes wave lines for 
gauge bosons; it is assumed that 
non-perturbative QCD completes the FS. Sometimes art helps somewhere. There is a real 
challenge, namely to connect amplitudes in the world of quarks with those in the world of hadrons that are and can be measured.  

\noindent 
This is a complex one on several levels. In the collisions of (anti-)baryons at low energies one hardly 
care about them being bound states of three constituent (anti-)quarks -- unless one describes their EDMs, 
where discuss the difference between current vs. constituent ones. When one talks about non-leptonic 
decays of hadrons, it is crucial to use current quarks. We know how to describe inclusive decays; 
however probing \cp\/\ot~violation the landscapes are much more complex, and we need more subtle tools to describe also multi-body FS. We can{\em not} focus only on two-body FS. 

\noindent 
There is a general comment: it is one thing to draw Feynman diagrams, but understanding 
the underlying forces is another thing; one needs more thinking and uses correlations with other 
transitions. One shows the impact of penguin diagrams in $K \to \pi \pi$ decays, although 
loop diagrams are usually suppressed. On the other hand their impact is enhanced by chiral symmetry 
for two pions FS and somewhat for three pions.  However this does not work 
for many-body FS in the decays of charm or beauty hadrons.

(12) Usually one compares the predictions from models with the information gotten by fitting best the data. There is a good reason to say that the analyses are model insensitive.  
However it is only the first (and second) step; in particular when one has a good candidate for a real theory, one has to 
focus whether these predictions come around within two sigma or so and think \& probe 
{\em correlations} with other data. Theoretical uncertainties are systematic at best; often `predictions' follow the fashion.   

\noindent 
It shows the connection of \CPV~with the violation of \ot~reversal (\TV). One might say it is  
obvious in 
$e^{i\phi t}$ to reach the same goal going down a different road. 
On the other hand, the landscape of \ot~reversal is very complex. It depends on its definition. 
For example, we know that it happens already in classical physics: it is much easier to 
get `down' than `up' -- i.e., the different scenarios of initial and final states. In this article I 
discuss \TV~in fundamental forces.

(13) There is a short comment about `oscillations' vs. `mixing'. Of course `mixing' covers 
more items in dynamics than `oscillation'. However I see no reason to use the same 
word for different regions of dynamics. 
\begin{itemize}
\item
`Oscillations' needs forces that can change the `flavor' by two units. 
Their impact 
depends on the time of decay in well-known and measurable ways.
\item
It depends on the initial neutral decaying hadrons like $K^0$ or $B_d$ \& $B_s$. 

\item
Oscillation is a much more  narrow meaning by focus on indirect \CPV; oscillation is a crucial step to probe \CPV,  
but it can happen without \CPV~ -- as we know so far about $D^0$ decays. 

\item
I prefer to use the word of `mixing' in narrow situations like $s \leftrightarrow d$ about the 
Cabibbo angle or in general: 
\begin{itemize}
\item 
It shows the connection of quarks with mass states that couple to weak charged bosons 
leading to the CKM matrix. 

\item
Likewise for `massive' neutrinos: they couple with charged leptons leading to the 
Pontecorvo--Maki--Nakagawa--Sakata matrix \cite{PMNS} (and maybe with Dark Matter).

\end{itemize}

\end{itemize}

\section{\cp~asymmetries in hadrons' decays}
\label{LAND}

\CPV~in neutral kaons and $B$ mesons have been established; 
it  depends on our quantitative understanding {\em quark flavor} dynamics including non-perturbative QCD. 
None have been found yet in charm hadrons; so far we have not enough rate to probe for top quarks \cite{CPBOOK}.

One first focuses on the transitions of neutral mesons with richer landscapes. \CPV~was found 
in $K_L$ or $\bar K^0/K^0$ and $\bar B^0/B^0$ decays. We have many examples, and the future data will 
show more. We use tools based on quantum mechanics \& quantum field theories and measure the correlations between different FS and differentiate
 the impact of ND and its features sometimes in subtle ways. It needs much more work \& analysis. However I will emphasize the informations we get from {\em many-body} 
 FS about direct \CPV~with accuracy and will discuss that about baryon and charged mesons decays below. 

$\Delta F \neq 0$ forces connect {\em neutral} flavor mesons $P^0$
 ($P^0=K^0,B_d,B_s,D^0$)  and $\bar P^0$. Therefore mass eigenstates are described linear amplitudes based on \cpt~\cite{CPBOOK}: 
\bea
|P_1\rangle = p |P^0 \rangle + q |\bar P^0\rangle \\
|P_2\rangle = p |P^0 \rangle - q |\bar P^0\rangle
\eea
are mass \& width eigenstates with eigenvalues \& their differences 
\footnote{There are opposite signs of $q/p$; using negative sign is equivalent to interchanging labels $1\leftrightarrow 2$.} : 
\bea
M_1 - \frac{i}{2}\Gamma_1 &=& M_{11} - \frac{i}{2}\Gamma_{11} + \frac{q}{p}(M_{12} - \frac{i}{2} \Gamma_{12}  ) \\
M_2 - \frac{i}{2}\Gamma_2 &=& M_{11} - \frac{i}{2}\Gamma_{11} - \frac{q}{p}(M_{12} - \frac{i}{2} \Gamma_{12}  ) \\     
M_2 - M_1 = -2 {\rm Re} \left(   \frac{q}{p}(M_{12} - \frac{i}{2} \Gamma_{12} )    \right)  &,&
\Gamma_2 - \Gamma_1 = + 4 {\rm Im} \left(   \frac{q}{p}(M_{12} - \frac{i}{2} \Gamma_{12} )   \right) \\
\left( \frac{q}{p}  \right) ^2= \frac{M_{12}^* - \frac{i}{2}\Gamma_{12}^* }{M_{12} - \frac{i}{2}\Gamma_{12}} &,& 
\frac{q}{p} = \sqrt{\frac{ M_{12}^* - \frac{i}{2}\Gamma_{12}^* }{ M_{12} - \frac{i}{2}\Gamma_{12}}  }
\eea
$q/p$ itself is {\em not} an observable. One can change the phase of anti-particles: 
$|\bar P^0\rangle \to e^{i\xi} |\bar P^0 \rangle$ will modify the off-diagonal elements of 
${\bf M}$ \& ${\bf \Gamma}$ and thus $q/p \to e^{-i\xi}q/p$. However 
both $|q/p|$ and $\frac{q}{p} (M_{12} - \frac{i}{2}\Gamma_{12} )$ are invariant and observable 
in different ways:  
\begin{itemize}
\item 
$P_1$ and $P_2$ states in general are not orthogonal to each other: 
$\langle P_1|P_2\rangle = |p|^2 - |q|^2 \neq 0$. 

\item 
This situation can be measured in semi-leptonic rates using \cpt\ invariance with 
$|A|^2 = |A(l^+)|^2 = |\bar A(l^-)|^2$: 
\bea
\Gamma (P^0 \to l^- +X^+) &\propto& e^{-\Gamma t} \left|\frac{q}{p}\right|^2 |A|^2 \\
\Gamma (\bar P^0 \to l^+ +X^-) &\propto& e^{-\Gamma t} \left|\frac{p}{q}\right|^2 |A|^2
\eea
One probes \CPV~based on $P^0 - \bar P^0$ oscillations, however it is independent of time:
\beq
A_{SL}(P^0) \equiv \frac{\Gamma (P^0 \to l^- +X^+)- \Gamma (\bar P^0 \to l^+ +X^-)}{\Gamma (P^0 \to l^- +X^+)+\Gamma (\bar P^0 \to l^+ +X^-) }
= \frac{1-|p/q|^2 }{1+|p/q|^2 }
\label{ASL}
\eeq 
It depends on the initial state with only indirect \CPV~with $\Delta F=2$, namely $K^0$, $B_d$, $B_s$ 
and $D^0$ transitions. 

\item 
Basically quantum mechanics tell us about the two mass eigenstates $P_1$ \& $P_2$ using the Schwartz inequality \cite{SCHINEQ,CPBOOK} arrive at 
\beq
|\langle P_2|P_1 \rangle | \leq 
\sqrt{\frac{\sum_f 4\Gamma_1^f\Gamma_2^f}{(\Gamma_1+\Gamma_2)^2+4(M_1-M_2)^2 }    }
\eeq
This inequality is numerically relevant for kaons due to $\Gamma_L \ll \Gamma_S \simeq 2\Delta M_K$:
\beq
|\langle K_L | K_S \rangle| \leq \sqrt{\frac{2\Gamma_L}{\Gamma_S} } \sim 0.06
\eeq
as a very conservative bound: $K_L$ and $K_S$ are close to be {\em odd} and 
{\em even} \cp\ states. 
Does it mean that we are just being lucky with $3m_{\pi} < M_K < 4m_{\pi}$  or is a deep reason 
about these limits?   

\item
The landscape is more complex with non-leptonic decays about indirect and direct \CPV~even with  
$f \neq \bar f$ based on quantum field theories with \cpt~invariance \cite{CPBOOK}: 
\bea
&&\Gamma (P^0 (t) \to f) \propto \frac{1}{2} e^{-\Gamma_1 t}|A(f)|^2 \cdot G_f(t) \\
G_f(t) &=& a+be^{\Delta \Gamma t } +ce^{\Delta \Gamma t/2}{\rm cos}\Delta M t +de^{\Delta \Gamma t/2} {\rm sin}\Delta M t \\
a&=&  \frac{1}{2}\left( 1+\left| \frac{q}{p}\bar \rho (f)  \right|^2   \right)+
{\rm Re}\left(\frac{q}{p}\bar \rho (f) \right) 
\\
b&=& \frac{1}{2} \left( 1+\left| \frac{q}{p}\bar \rho (f)  \right|^2 \right)-
{\rm Re}\left(\frac{q}{p}\bar \rho (f) \right) 
\\
c&=&1- \left| \frac{q}{p}\bar \rho (f)  \right|^2 \; , \; 
d= -2{\rm Im}\frac{q}{p}\bar \rho (f) \\
&&\Gamma (\bar P^0 (t) \to \bar f) \propto \frac{1}{2} e^{-\Gamma_1 t}|\bar A( \bar f)|^2 \cdot 
\bar G_{\bar f}(t) \\
\bar G_{\bar f}(t) &=& \bar a+\bar b e^{\Delta \Gamma t } 
+\bar c e^{\Delta \Gamma t/2}{\rm cos}
\Delta M t +
\bar de^{\Delta \Gamma t/2} {\rm sin}\Delta M t \\
\bar a&=&  \frac{1}{2} \left( 1+\left| \frac{p}{q} \rho (\bar f) \right|^2 \right)+
{\rm Re} \left(\frac{p}{q} \rho (\bar f) \right) 
\\
\bar b&=& \frac{1}{2}\left( 1+\left| \frac{p}{q} \rho (\bar f)   \right|^2 \right )-
{\rm Re}\left(\frac{p}{q} \rho (\bar f) \right) 
\\
\bar c&=&1- \left| \frac{p}{q} \rho (\bar f)  \right|^2 \; , \; 
\bar d= -2{\rm Im}\frac{p}{q} \rho (\bar f)
\eea
\beq
\bar \rho (f) = \frac{\bar A(f)}{A(f)}\; , \;
\rho (\bar f) = \frac{A(\bar f)}{\bar A(\bar f)} 
\eeq 
It is important that $(q/p)\bar \rho (f)$ and $(p/q) \rho (\bar f)$ do {\em not} depend 
on the definition of the phases and therefore are observables, while $(q/p)$ and $\bar \rho (f)$ by itself are not. 
 
\end{itemize}
Here you can see the general situation. In our world we have $|q/p| \sim 1$ and 
$\Delta \Gamma/\Gamma \sim 0$ (except for 
$\Delta \Gamma (K_L)/(\Gamma(K_L)+\Gamma (K_S)) \simeq \Delta \Gamma (K_L)/\Gamma (K_S)  \simeq 0.49$). 
 $\Delta \Gamma = 0$ can happen only due to a miracle; however, $\Delta \Gamma$ can be smaller 
or larger than expected from SM values; the impact of ND can hide in the experimental 
and/or theoretical uncertainties. 

When the FS are even/odd \cp~eigenstates, one gets
\beq 
\bar \rho (f_{\pm}) = \pm \frac{1}{\rho (f_{\pm})} \; .  
\eeq

For {\em charged} $P$ mesons the landscapes look much simpler 
\footnote{One can easily connect the expressions given for $P^0$ vs. $\bar P^0$ with 
$\Delta \Gamma = 0 = \Delta M$.}
-- but not very much in reality:
\bea
\Gamma (P \to f_a) &\propto& e^{-\Gamma t} |A(f_a)|^2 \\
\Gamma (\bar P \to \bar f_a) &\propto& e^{-\Gamma t} |\bar A( \bar f_a)|^2 
\eea
There are several important statements, although they are not always obvious:
\begin{itemize}
\item 
Time depending rates show the impact of indirect vs. direct \CPV\ in neutral mesons. 
The amplitudes of indirect ones depend in the initial states: $K^0$, $B_d$, $B_s$ and $D^0$. 
These can be probed in two-body FS. 

\item 
Direct \CPV~affect differently FS in the decays of hadrons. 
It is not enough to understand the dynamics with two-body FS; it is crucial 
to measure three- and four-body FS with accuracy, but not only as a back-up information.

\item 
The impact of strong re-scattering is crucial as discussed below. It  happens in the 
world of hadrons (and of quarks) as indicated in $f_a$ vs. $\bar f_a$; however it is a true challenge to describe them 
quantitatively with subtle theoretical tools. 

\end{itemize}

\section{The landscapes of strange, beauty \& charm hadrons} 
\label{LANDGENCPV}

\subsection{Kaon decays -- first `affair'}
\label{KAON1}

The existence \CPV~was first found in 1964 by  $K_L\to \pi^+\pi^-$ -- 
i.e., the $K_L$ amplitude 
has a small non-zero \cp~odd component due to $K^0 - \bar K^0$ oscillations with 
$\Delta M_K/\Gamma_S \simeq 0.49$: 
\beq
\frac{\Gamma(K_L \to \pi^+\pi^-)}{\Gamma(K_S \to \pi^+\pi^-)} = [(2.0 \pm 0.4)] \times 10^{-3}]^2
\eeq 
The existence of this small rate is connected with the asymmetry in 
$\bar K^0 \to l^+\nu \pi^-$ vs. $K^0 \to l^- \bar \nu \pi^+$ in the SM (and basically beyond) --  i.e., indirect \CPV:
\beq
A_L = \frac{\Gamma(K_L \to l^+\nu_l\pi^-)-\Gamma(K_L\to l^-\bar \nu \pi^+)}
{\Gamma(K_L \to l^+\nu_l\pi^-)+\Gamma(K_L\to l^-\bar \nu \pi^+)} 
= (3.32 \pm 0.06) \cdot 10^{-3} \; .
\eeq
While $A_L$ comes from oscillations, this asymmetry does not depend on the time of the decays. 
The scenarios of non-leptonic decays are more complex also for $K_L$ decays: weak forces 
produce $K_L \to \pi^+\pi^-/\pi^0\pi^0$, which are calibrated by $K_S$ decays: 
\beq
\eta_{+-} \equiv \frac{\langle \pi^+\pi^- |H_W|K_L \rangle }{\langle \pi^+\pi^- |H_W|K_S \rangle} \; , \; 
\eta_{00} \equiv \frac{\langle \pi^0\pi^0 |H_W|K_L \rangle }{\langle \pi^0\pi^0 |H_W|K_S \rangle}
\eeq
We differentiate indirect vs. direct \CPV: 
\beq
\eta_{+-} = \epsilon_K +\epsilon^{\prime} \; , \; \eta_{00} = \epsilon_K -2\epsilon^{\prime} \; , 
\eeq
where $\epsilon_K$ is produced by oscillations, while $\epsilon^{\prime}$ show the differences between 
different FS. Present data show \cite{PDG}:
\beq
|\epsilon_K| = (2.228 \pm 0.011) \cdot 10^{-3} 
\eeq 
The response from the theoretical community about \CPV~was slow. It was suggested by Wolfenstein that we 
have a ND, namely super-weak one with $\epsilon^{\prime}=0$. However there was not a real theory, 
but a classification of theories for \CPV. 

\subsection{New Standard Model `then'}
\label{KMTHEORY}

Most HEP people `knew' about three quarks, namely $u,d,s$; most thought of them as a mathemalical  
entities. Some outliers told about the fourth quark, namely $c$. To understand to underlying dynamics 
of \CPV~Kobayashi \& Maskawa published a paper in 1973 that there are three classes of theories beyond the 
SM then, namely at least three families of quarks or right-handed charged currents or charged Higgs. 
Now we know that at least the leading source comes from three families with $(u,d)$, $(c,s)$ and 
$(t,b)$ with weak forces $SU(2)_L \times U(1)$. 

We have to deal with somewhat different landscapes, namely we can probe data based on 
hadrons and predict transitions based on quantum field theories with quarks \& spin-one weak bosons \& gluons.  
This connection comes from the word of `duality' in different levels; some are obvious, others are subtle.  

For the SM one gets an unitary
CKM matrix for three families with six charged quarks in pairs $(u,d)$, $(c,s)$ and $(t,b)$: 
\begin{eqnarray}
V_{\rm CKM}= \left(\footnotesize \begin{array}{ccc} 
V_{ud} & V_{us} & V_{ub} \\
V_{cd} & V_{cs} & V_{cb} \\
V_{td} & V_{ts} & V_{tb}
\end{array}\right)  
\end{eqnarray}
It is described by six triangles. 
However one gets only four observables, namely three angles and one phase. 
Their patterns are quite different, but their have the same area. 
The general parameterization of flavor dynamics is not obvious.

\subsubsection{Wolfenstein's original parameterization \& refined ones}
\label{WEAK}

Wolfenstein suggested a very good `usable' one based on the expansion in the 
Cabibbo angle $\lambda = {\rm sin}\theta_C \simeq 0.223$ with the
 three $A$, $\rho$ and $\eta$ being 
of the order of unity \cite{WOLFMAT}: 
\begin{eqnarray} 
{\bf {\rm V}}_{\rm CKM} \simeq 
\left(\footnotesize
\begin{array}{ccc}
 1 - \frac{\lambda ^2}{2}  & \lambda , & 
 A\lambda^3(\rho - i\eta)  \\
-\lambda & 1 - \frac{\lambda ^2}{2} - i\eta A^2\lambda^4 & 
A\lambda^2(1+i\eta\lambda^2)
\\
A\lambda^3(1-\rho - i\eta) & -A\lambda^2 &1
\end{array}
\right)
\end{eqnarray}
\bea
{\rm `Old' \, triangle\; I.1:}&&V_{ud}V^*_{us} \; \; \;  [{\cal O}(\lambda )] + V_{cd}V^*_{cs} \;  \; \;  
[{\cal O}(\lambda )] + 
 V_{td}V^*_{ts} \; \; \; [{\cal O}(\lambda ^{5} )] = 0   \\ 
{\rm `Old' \, triangle\; I.2:}&& V^*_{ud}V_{cd} \; \; \;  [{\cal O}(\lambda )] + V^*_{us}V_{cs} \; \; \;  [{\cal O}(\lambda )] + 
V^*_{ub}V^*_{cb} \; \; \; [{\cal O}(\lambda ^{5} )] = 0    \\
{\rm `Old' \, triangle\; II.1:}&& V_{us}V^*_{ub} \; \; \;  [{\cal O}(\lambda ^4)] + V_{cs}V^*_{cb} \;  \; \;  [{\cal O}(\lambda ^{2} )] + 
V_{ts}V^*_{tb} \; \; \; [{\cal O}(\lambda ^2  )] = 0   \\ 
{\rm `Old' \, triangle\; II.2:}&& V^*_{cd}V_{td} \; \; \;  [{\cal O}(\lambda ^4 )] + V^*_{cs}V_{ts} \; \; \;  [{\cal O}(\lambda ^{2})] + 
V^*_{cb}V^*_{tb} \; \; \; [{\cal O}(\lambda ^{2} )] = 0  \\
{\rm `Old'\, triangle\; III.1:}&& V_{ud}V^*_{ub} \; \; \;  [{\cal O}(\lambda ^3)] + V_{cd}V^*_{cb} \;  \; \;  [{\cal O}(\lambda ^{3} )] + 
V_{td}V^*_{tb} \; \; \; [{\cal O}(\lambda ^3  )] = 0   \\ 
{\rm `Old' \, triangle\; III.2:}&& V^*_{ud}V_{td} \; \; \;  [{\cal O}(\lambda ^3 )] + V^*_{us}V_{ts} \; \; \;  
[{\cal O}(\lambda ^{3})] + 
V^*_{ub}V^*_{tb} \; \; \; [{\cal O}(\lambda ^3 )] = 0 
\label{NEW2}
\eea 
\begin{figure}[h!]
\begin{center}
\includegraphics[width=6cm]{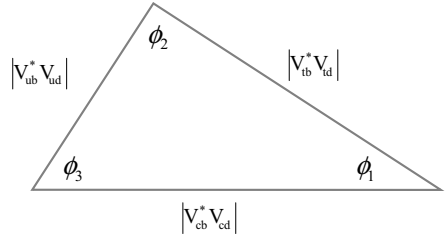}
\end{center}
\caption{The `golden' CKM unitarity triangle.}
\label{fig:CKMTRIANGLENOT}
\end{figure}
\begin{figure}[h!]
\begin{center}
\includegraphics[width=6cm]{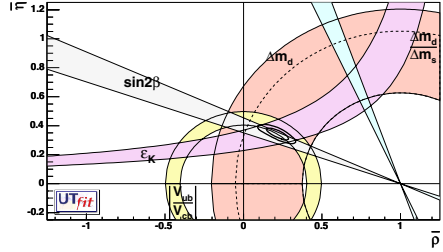}
\end{center}
\caption{The `golden' CKM unitarity triangle fitted including the impacts from $\epsilon_K$ 
and $\Delta M_{B_s}$ from two other triangles.}
\label{fig:CKMTRIANGLEFIT}
\end{figure}
The pattern is obvious, in particular about indirect \CPV~, namely very larger in $B_d-\bar B_d$ oscillations. 
It has been successful in describing the `golden' triangle in 
$B_{d,u}$ decays in Fig.\ref{fig:CKMTRIANGLENOT}, where triangle III.1 shows that the sizes of the three angles are quite similar. 
The angles $\phi_1$, $\phi_2$, $\phi_3$ are opposite the 
sides with $\bar u u$, $\bar cc$, $\bar t t$; other people name angles $\beta$, $\alpha$, 
$\gamma$. 
 
It is crucial to probe the correlations in the triangles. 
Fig.\ref{fig:CKMTRIANGLEFIT} shows that large \CPV~in $B_d \to \psi K_S$ is connected with 
very small one in $K_L \to \pi \pi$ transitions and the ratio of $B_d-\bar B_d$ \& 
$B_s-\bar B_s$ oscillations due to $\Delta M_{B_d}/\Delta M_{B_s}$; i.e., those observables (mostly) come from three triangles.

That successful description still has some weak points. There are some `tensions' about the  
data and the expected predictions. 
Furthermore measured decays of $B_{d,u}$ mesons give us $\eta \simeq 0.34$ and 
$\rho \simeq 0.13$, which are not close to unity.  
With three families of quarks one gets six triangles to decays of kaons, charm \& 
beauty decays and top quarks. Four of those six ones one can probe directly. 
The patterns of these triangles are very different.

The SM produces at least the leading source of \CPV~in $K_L \to 2 \pi$ and 
$B$ decays with good accuracy. Searching for ND we need even precision and to measure  
the correlations with other FS's. The landscape of the CKM matrix is more subtle as pointed out \cite{AHN} 
\begin{eqnarray} 
\left(\footnotesize
\begin{array}{ccc}
 1 - \frac{\lambda ^2}{2} - \frac{\lambda ^4}{8} - \frac{\lambda ^6}{16}, & \lambda , & 
 \bar h\lambda ^4 e^{-i\delta_{\rm QM}} , \\
 &&\\
 - \lambda + \frac{\lambda ^5}{2} f^2,  & 
 1 - \frac{\lambda ^2}{2}- \frac{\lambda ^4}{8}(1+ 4f^2) 
 -f \bar h \lambda^5e^{i\delta_{\rm QM}}  &
   f \lambda ^2 +  \bar h\lambda ^3 e^{-i\delta_{\rm QM}}   \\
    & +\frac{\lambda^6}{16}(4f^2 - 4 \bar h^2 -1  ) ,& -  \frac{\lambda ^5}{2} \bar h e^{-i\delta_{\rm QM}}, \\
    &&\\
 f \lambda ^3 ,&  
 -f \lambda ^2 -  \bar h\lambda ^3 e^{i\delta_{\rm QM}}  & 
 1 - \frac{\lambda ^4}{2} f^2 -f \bar h\lambda ^5 e^{-i\delta_{\rm QM}}  \\
 & +  \frac{\lambda ^4}{2} f + \frac{\lambda ^6}{8} f  ,
  &  -  \frac{\lambda ^6}{2}\bar h^2  \\
\end{array}
\right)
+ {\cal O}(\lambda ^7)
\end{eqnarray}
with $\bar h \simeq 1.35$, $f\simeq 0.75$ \& $\delta _{\rm QM} \sim 90^{o}$ and only  
expansion in $\lambda \simeq 0.223$.  

The `shapes' of six triangles are 
different in subtle ways:  
\bea
{\rm Triangle\; I.1:}&&V_{ud}V^*_{us} \; \; \;  [{\cal O}(\lambda )] + V_{cd}V^*_{cs} \;  \; \;  [{\cal O}(\lambda )] + 
 V_{td}V^*_{ts} \; \; \; [{\cal O}(\lambda ^{5\& 6} )] = 0   \\ 
{\rm Triangle\; I.2:}&& V^*_{ud}V_{cd} \; \; \;  [{\cal O}(\lambda )] + V^*_{us}V_{cs} \; \; \;  [{\cal O}(\lambda )] + 
V^*_{ub}V^*_{cb} \; \; \; [{\cal O}(\lambda ^{6 \& 7} )] = 0    \\
{\rm Triangle\; II.1:}&& V_{us}V^*_{ub} \; \; \;  [{\cal O}(\lambda ^5)] + V_{cs}V^*_{cb} \;  \; \;  [{\cal O}(\lambda ^{2 \& 3} )] + 
V_{ts}V^*_{tb} \; \; \; [{\cal O}(\lambda ^2  )] = 0   \\ 
{\rm Triangle\; II.2:}&& V^*_{cd}V_{td} \; \; \;  [{\cal O}(\lambda ^4 )] + V^*_{cs}V_{ts} \; \; \;  [{\cal O}(\lambda ^{2\& 3})] + 
V^*_{cb}V^*_{tb} \; \; \; [{\cal O}(\lambda ^{2 \& 3} )] = 0  \\
{\rm Triangle\; III.1:}&& V_{ud}V^*_{ub} \; \; \;  [{\cal O}(\lambda ^4)] + V_{cd}V^*_{cb} \;  \; \;  [{\cal O}(\lambda ^{3\& 4} )] + 
V_{td}V^*_{tb} \; \; \; [{\cal O}(\lambda ^3  )] = 0   \\ 
{\rm Triangle\; III.2:}&& V^*_{ud}V_{td} \; \; \;  [{\cal O}(\lambda ^3 )] + V^*_{us}V_{ts} \; \; \;  
[{\cal O}(\lambda ^{3\& 4})] + 
V^*_{ub}V^*_{tb} \; \; \; [{\cal O}(\lambda ^4 )] = 0 
\label{NEW2}
\eea  
The pattern in flavour dynamics is less obvious for \CPV~in hadron decays as stated before \cite{BUZIOZ}; 
the situation has changed: we have to measure the correlations between four triangles, not focus only on the 
`golden triangle'.  
Some of the important points are emphasized: 
\begin{itemize}
\item 
We have to probe triangle III.1 with precision in $B_{d,u}$ transitions. 
\item 
Triangle II.1 has sizable impact on $B_s$ amplitudes and connect with other $B_{d,u}$ decays. 

\item
Triangle I.2 produces \cp~asymmetries in SCS  $D$ decays, but hardly in DCS ones. 

\item
Triangle I.1 can be probed in tiny $K \to \pi \nu \bar \nu$ decays with small {\em theoretical} uncertainties. 
\item 
Again: one has to focus on {\em correlations} with several triangles with 
accuracy \footnote{I see a connection of `correlations' for a well-known joke: "In a circus an artist put tables 
\& chairs together and jump to the top with a head-stand using broom-stick to 
produce balance -- and play with a fiddle. One of the men watching that said to his wife: 
He is not like Haifetz."}.

\end{itemize}

\subsection{Kaon decays -- second `affair'}
\label{KAON2}

The measured values of $|\epsilon_K|$ gives small experimental uncertainty; the challenge is to connect 
with the CKM parameters as shown in Fig.\ref{fig:CKMTRIANGLEFIT}, namely mostly the impact of 
triangle I.1 on the golden one in triangle III.1. 

Direct \CPV~is expressed through the ratio:
\beq
{\rm Re}\frac{\epsilon^{\prime}}{\epsilon_K} = \frac{1}{6} \frac{|\eta_{+-}|^2 - |\eta_{00}|^2}{|\eta_{+-}|^2} = (1.66 \pm 0.23) \cdot 10^{-3}
\eeq  
These values do not do justice to the experimental achievement. The achievement becomes  
more transparent \cite{CPBOOK}: 
\beq
\frac{\Gamma (K^0 \to \pi^+\pi^-) - \Gamma (\bar K^0 \to \pi^+\pi^-) }
{\Gamma (K^0 \to \pi^+\pi^-) + \Gamma (\bar K^0 \to \pi^+\pi^-) } = (5.16 \pm 0.71)\cdot 10^{-6}
\eeq
There is no surprise that Re$(\epsilon^{\prime}/\epsilon_K)$ is small with the large top quark mass including the impact of penguin diagrams, see Fig.\ref{fig:nicepeng}. 
However the experimental uncertainty is sizable, and now Re$(\epsilon^{\prime}/\epsilon_K)$ 
gives no constraint on CKM parameters. More data and refined 
analyses of $K \to \pi \gamma \gamma$ \& $K \to \pi \pi \gamma \gamma$ allow deeper 
probes of chiral symmetry and in general better treatment of long-distance dynamics. It gives 
tests of LQCD as a subtle tool. Furthermore the LQCD community might be able to show that 
Re$(\epsilon^{\prime}/\epsilon_K)$ gives sizable (\& novel) impact on the correlations with 
the golden triangle \cite{LEHNER}. 

\subsection{Future of very rare kaon decays}
\label{KAON3}

There is still an important point about understanding fundamental dynamics, namely to 
measure the rates of  $K^+ \to \pi^+ \bar \nu \nu$ vs. $K_L \to \pi^0 \bar \nu \nu$ 
\bea
{\rm BR}(K^+ \to \pi^+ \bar \nu \nu) =(8.4\pm 1.0) \cdot 10^{-11} \\
{\rm BR}(K_L \to \pi^0 \bar \nu \nu) =(3.4\pm 0.6) \cdot 10^{-11}
\label{RAREKAON}
\eea
and probe $V_{td}$ with SM prediction with 5\% vs. 2\% theoretical uncertainties, respectively 
\cite{BURAS}. 
The only challenge is to get enough data with refined analyses. Thus they could act as "standard candles" in the future -- maybe.

\subsection{\cp~asymmetry in the decays of charged mesons \& baryons}
\label{ONLYDIR}

While `only' direct \CPV~can effect the decays of baryons and charged mesons, one might 
think that the landscape is less complex. The opposite is (mostly) true: 

\noindent 
(i)  Direct \CPV~depend on FS and on the classes of decaying hadrons. 

\noindent 
(ii) \cp~asymmetries do not depend on the times of the decays. Measuring time depending 
asymmetries is a very powerful tool.

\noindent 
(iii) One has to focus even more importantly on {\em regional} \CPV~where 
one needs at least three pseudo-scalar ones in the FS. 
As discussed below large ones have been found in $B^{\pm}$.

\subsection{Effective transition amplitudes}
\label{EFFECT}
 
Strong re-scatterings happen all the time. Can we can describe them quantitatively?  
Our control of non-perturbative QCD is limited so far. However it helps to deal with this challenge with tools following  constraints 
coming from symmetries (broken or not). It has large impact on direct \cp~asymmetries 
with \cpt~invariance as discussed in Refs.\cite{1988BOOK},\cite{FSICP},\cite{WOLFFSI}; 
it is given in Sect. 4.10 of Ref.\cite{CPBOOK} with much more details:
\bea
T(P \to f) &=& e^{i\delta_f} \left[ T_f + 
\sum_{f \neq a_j}T_{a_j}iT^{\rm resc}_{a_jf}  \right] 
\label{CPTAMP1} 
\\
T(\bar P \to \bar f) &=& e^{i\delta_f} \left[ T^*_f +
\sum_{f \neq a_j}T^*_{a_j}iT^{\rm resc}_{a_jf}  \right]   \; , 
\label{CPTAMP2}
\eea 
where amplitudes $T^{\rm resc}_{a_jf}$ describe FSI between $f$ and intermediate 
{\em on}-shell states $a_j$ that connect with this FS. Thus one gets {\em regional} 
\cp~asymmetries: 
\beq
\Delta \gamma (f) = |T(\bar P \to \bar f)|^2 - |T(P \to f)|^2 = 
4 \sum_{f \neq a_j} T^{\rm resc}_{a_jf} \, {\rm Im} T^*_f T_{a_j} 
\label{REGCPV}
\eeq
\cp~asymmetries have to vanish upon summing over all such states $f$ 
using \cpt~invariance between {\em subclasses} of partial widths: 
\beq
\sum_{f} \Delta \gamma (f) =
4 \sum_{f}\sum_{f \neq a_j} T^{\rm resc}_{a_jf} {\rm Im} T_f^* T_{a_j} = 0 \; , 
\eeq
since $T^{\rm resc}_{a_jf}$ \& Im$T_f^* T_{a_j}$ are symmetric \& antisymmetric, 
respectively, in the indices $f$ \& $a_j$. 

These Eqs. (\ref{CPTAMP1},\ref{CPTAMP2}) apply to amplitudes in general, whether for hadrons or quarks  
or $\bar q_i q_j$ boundstates in between  
\footnote{In principle one has to include baryons $q_iq_jq_k$, but I will not discuss that in this article.}. 
In which way one can connect the landscapes in hadronic and quark amplitudes -- it depends. 
In the world of quarks one can describe them by refined tree, penguin etc. diagrams.  Those give 
weak phases. Furthermore penguin diagrams coming from non-local operators  
produce $\Delta \Gamma$ for $B_{s,d}$ mesons and somewhat for $D^0$ one. Those give 
also imaginary part that one needs for FSI -- however the situations are very `complex' there. 
The `roads' are quite different depending on the FS.

\subsection{Impact of non-perturbative QCD}
\label{QCD}

The scenarios of the weak decays of beauty and charm hadrons are more complex than in kaon ones. 
There are several reasons:
\begin{itemize}
\item 
Two-body FS produce only small parts of CKM suppressed of $D_{(s)}$ decays and tiny  
in $B_{(s)}$ ones. 
There is no reason why two-body  FS give us all the information that we need to understand 
dynamics and even less for only charged two-body ones.  

\item 
The worlds of hadrons and quarks are different. One can hide that by using `constitute'  
quarks, which works fine for spectroscopy (in particular for strange hadrons), but 
not for weak forces. `Current' quarks are based on theories, not just models. However they are 
connected in subtle ways, and we have to apply refined tools.

\item
In the world of quarks one can describe inclusive FS in beauty \& charm hadrons, 
where we have to use `duality' often in subtle ways. 
\cpt~invariance produces strong constraints. To connect finite data of hadrons with quarks descriptions 
one has to use tools based on chiral symmetry, broken U-spin symmetry, 
dispersion relations \cite{DISPERSION} etc. and  insist on correlations with other transitions. 

\item 
Probing \CPV~in many-body FS one measures first {\em averaged} one and then {\em regional} ones 
with accuracies. It is not a good idea to just follow the best fits; it is much more important to understand 
the landscapes and their informations given to us. 
Of course the analysis has to be very acceptable -- but not giving the best fits. Judgment helps 
significantly how to define regional asymmetries.  

\item 
One measures three-body FS for several reasons with a long history \cite{GOLDEN}. 
We know how to probe Dalitz plots including regional `morphologies'; it has been emphasized not only use 
`fractional' asymmetries, but also about different tools \cite{MIRANDA} and compare their results. 
In my view 
it is not  the final step; we have to use more sutble theoretical tools like dispersion relations 
that depend also on data about low energy collisions of hadrons \cite{DISPERSION} -- and 
some judgment.

\end{itemize}
The landscapes are very different already qualitatively between $\Delta B \neq 0$ and 
$\Delta C\neq 0$.

\subsubsection{Case I: Broken U-spin symmetry}

With quarks one describes mostly {\em inclusive} transitions. `Currents'  
quarks with $m_u < m_d \ll m_s$ are based on theory. I-, U- \& V-spin symmetries deal 
with $u \leftrightarrow d$, $d \leftrightarrow s$ \& $u \leftrightarrow s$. 
These three symmetries are obviously broken on different levels, and these violations are  
connected in the SM. The operators producing inclusive FS depend on their CKM 
parameters and the current quark masses involved there. However the real scale for inclusive decays is given by 
the impact of QCD, namely $\bar \Lambda \sim  1$ GeV as discussed many times 
\footnote{For good reasons one uses different and smaller $\Lambda_{\rm QCD} \sim 0.1 - 0.3$ GeV for describing jets in collisions.}. 
Thus the violations of U- \& V-spin symmetries are small, and tiny for I-spin one. 
We can deal with inclusive rates and asymmetries of beauty and maybe charm hadrons using  
effective operators in the world of quarks. 

The connections of inclusive with exclusive hadronic rates are not obvious at least, in particular about quantitative ways. The violations of 
I-, U-(\& V-)spin symmetries in the measurable world of hadrons are expected to scale by the differences in pion and kaon masses, 
which are {\em not} small compared to $\bar \Lambda$ (or  
$[m^2_K - m^2_{\pi}]/[m^2_K + m^2_{\pi}]$). 
This is even more crucial about direct \CPV~and the impact of strong re-scattering on amplitudes. 

Going back to the history: Lipkin had suggested that U-spin violations in $B$ decays 
are of the order of 10 \% \cite{LIPK2} in CKM {\em favored} ones. They might be larger in {\em suppressed} ones. The worlds of 
hadrons (or `constitute' quarks) are controlled by FSI due to {\em non}-perturbative QCD; they show the stronger  impact on exclusive ones.
For good reasons it has been stated that violation of U-spin symmetry is
around ${\cal O}(10 \%)$ in inclusive decays. It can be seen in the sum of {\em exclusive} ones in 
large ratios that go up and down much more sizably. The papers \cite{BGR} suggest one 
can probe U-spin symmetry with three-body FS with small theoretical uncertainties and even with only 
{\em charged hadrons} in the FS; I quite disagree on both, see \cite{BUZIOZ,MANY,TIM} with more comments \& details: 
`Effective transition 
amplitudes' or re-scattering as discussed above (see Sect. \ref{EFFECT}) produce large impact. I suggest to think about the informations gotten from 
Sect. \ref{DIRBDECAY} using \cpt~invariance about their subtle 
morphologies discussed below.

\subsubsection{Case II: Impact of penguin operators vs. diagrams}
\label{PENG}

Penguin diagrams (Fig.\ref{fig:nicepeng}) were introduced for kaon decays where is little differences between exclusive vs. inclusive decays. The impact of penguin operators in 
CKM suppressed decays of beauty hadrons are enhanced by chiral symmetry in their amplitudes, 
in particular for two body FS with pions and somewhat for kaons. However in charm hadron transitions the leading source of penguin diagrams is {\em not} given by local or even short-distance dynamics . 

\subsection{$B_{u,d,s}$ decays}
\label{BEAUTY}

The SM gives at least the leading source of \cp~asymmetries in $B$ transitions 
[with the still possible exception in $B_s \to \psi \phi, \psi f_0(980)$ ones]. Now 
we are probing for impact of ND in \cp~asymmetries and its features. 

\subsubsection{Indirect \CPV~in $B^0 - \bar B^0$ oscillations}
\label{BOSC}

Using $\Delta \Gamma _{B_d} \ll \Delta M_{B_d} \sim \Gamma _{B_d}$ as expected due to the large 
top quark mass, one states:
\bea
\Gamma (B_d [\bar B_d] \to \psi K_S) &\propto& e^{-\Gamma_Bt}G_{\psi K_S}[\bar G_{\psi K_S}] \\
G_{\psi K_S} &=& |A(\psi K_S|^2 \left[ 1 - {\rm Im}\left(\frac{q}{p}\bar \rho (\psi K_S) 
{\rm sim} \Delta M_{B_d} t   \right)\right]  \\
\bar G_{\psi K_S} &=&|A(\psi K_S|^2\left[ 1 + {\rm Im}\left(\frac{q}{p}\bar \rho (\psi K_S) 
{\rm sim} \Delta M_{B_d} t   \right)\right] \; .
\eea
\CPV~ is described by ${\rm Im}\left(\frac{q}{p}\bar \rho (\psi K_S)\right)$. One first 
needs 
$\Delta M_{B_d} \neq 0$ to measure this asymmetry, which depends on the time of decay. 
In other words:   
$ \frac{d}{dt}(G_{\psi K_S}/\bar G_{\psi K_S}) \neq 0 $ actually in a special way: sin$\Delta M_{B_d}t$, 
shows the connection with `odd' \ot~symmetry 
\footnote{To be precise: PDG2015 gives a value for $B^0 \to J/\psi(nS)K^0$; I ignore direct CPV in those 
FS, while PDG2015 gives $C(B^0 \to J/\psi(nS)K^0) = (0.5 \pm 2.0)\cdot 10^{-2}$.}: 
\beq
{\rm Im}\left(\frac{q}{p}\bar \rho (\psi K_S)\right) = 0.676 \pm 0.021
\eeq
In the SM for the `golden' triangle one gets Im$\left(\frac{q}{p}\bar \rho (\psi K_S)\right)$ $\simeq$ 
sin$2\phi_1[\beta]$. 

Refined parameterization of the CKM matrix show that the maximal value possible in the SM 
is $\sim$ 0.72 \cite{BUZIOZ}, not really close to unity due to correlations with other transitions. 

The situation is different about $B_s - \bar B_s$ oscillations with $\Delta M_{B_s}\simeq 26.9$ 
and $y_s \sim 0.07$: very fast oscillation has been established, but no \CPV~has been 
found (yet): 
\beq
\phi_{\rm s}^{c \bar cs} = (0.01 \pm 0.07 \pm 0.01)\; {\rm rad} \; {\rm (measured)} \; \;
{\rm vs.} \; \; \phi_{\rm s}^{c \bar cs} = (-0.0363 ^{+0.0016}_{-0.0015}) \; {\rm rad} \; {\rm (SM)}
\eeq
These data are close to SM values, but also consistent with ND's sizable 
contributions -- even leading source there -- or with the opposite sign. It is interesting that 
recent LHCb data about $B_s \to \psi \pi^+\pi^- \Rightarrow \psi f(980)$ see no obvious 
contribution from scalar $\sigma \Rightarrow \pi^+\pi^-$.

\subsubsection{Direct \CPV~in $B$ decays}
\label{DIRBDECAY}

The situations of decays of $B_{u,d,s}$ (and even $B_c$) are complex (for 
optimistic physicists); they are `rich' where one can find the impact of ND or at least important lessons about non-perturbative forces from QCD. 
Again first one focus on (quasi-)two-body FS about sizable asymmetries in 
$B^+ \to D_{\rm CP+}K^+$, 
which has impact of measuring the angle $\phi_3/\gamma$. Furthermore penguin diagrams 
contribute to \CPV~in $B_d \to K^+\pi^-$, $K^*(892)^+\pi^-$, $B_s \to \pi^+K^-$ and 
$B^+ \to \eta K^+$ on different levels  
\footnote{Here one has also interference with indirect \CPV~in $B_d \to \pi^+\pi^-$.}. 
The real challenge is to establish the impact of ND as a non-leading source. 
In the world of quarks one can show the ways to connect with hadronic FS with 
penguin diagrams due `duality', which is a true challenge in a quantitative way. 

Probing \CPV~in the SM suppressed decays one gets only a number in two-body FS. 
To connect the information we get from the data with the fundamental dynamics is not trival -- 
but it is {\em not enough} about forces: we have to probe three- \& four-body FS etc. We describe three-body FS due to two-dimensional Dalitz plots. The first 
step is to measure averaged \CPV~which also give numbers, but still connected with two-body ones. However it is crucial to probe regional asymmetries. 
I give recent examples about the power and the tools including \cpt~invariance.

\subsubsection{\cp~asymmetries in $B^{\pm}$ decays}
\label{BPM}

In this article I focus on charged three-body FS, although I will talk also about the  
general landscape including \cpt. LHCb data give small rates for 
CKM suppressed $B^+$ decays to charged three-body FS, which are not unusual:  
\bea
\nonumber 
{\rm BR}(B^+ \to K^+\pi^-\pi^+) = (5.10 \pm 0.29 ) \cdot 10^{-5}   \\   
{\rm BR}(B^+ \to K^+K^-K^+) = (3.37 \pm 0.22 ) \cdot 10^{-5} 
\label{KPIPI}
\eea
LHCb data also show sizable \cp~asymmetries {\em averaged} over the FS \cite{LHCb028}: 
\bea
\Delta A_{CP}(B^{\pm} \to K^{\pm} \pi^+\pi^-) &=&  
+0.032 \pm 0.008_{\rm stat} \pm 0.004_{\rm syst}
[\pm 0.007_{\psi K^{\pm}}]  
\nonumber \\
\Delta A_{CP}(B^{\pm} \to K^{\pm} K^+K^-) &=&   
- 0.043 \pm 0.009_{\rm stat} \pm 0.003_{\rm syst}
[\pm 0.007_{\psi K^{\pm}}] \; .
\label{SUPP1}
\eea 
It is very interesting that  these \cp~asymmetries come with {\em opposite} signs 
due to the road to \cpt~invariance and give us lessons about underlying dynamics. Still it is 
not surprising. Furthermore they show `regional' \cp~asymmetries {\em defined} by the LHCb collaboration: 
\bea 
A_{CP}(B^{\pm} \to K^{\pm} \pi^+\pi^-)|_{\rm regional} &=& + 0.678 \pm 0.078_{\rm stat} 
\pm 0.032_{\rm syst}
[\pm 0.007_{\psi K^{\pm}}] 
\nonumber \\
A_{CP}(B^{\pm} \to K^{\pm} K^+K^-)|_{\rm regional} &=& - 0.226 \pm 0.020_{\rm stat} \pm 
0.004_{\rm syst}
[\pm 0.007_{\psi K^{\pm}}] \; .
\label{SUPP3}
\eea 
One expects that `regional' asymmetries are larger than averaged ones. At least they show the impact of re-scattering. Again, one sees the opposite signs; 
however the sizes are quite different. Furthermore one 
has to remember that scalar resonances (like $f_0(500)/\sigma$ \& $\kappa$) produce broad ones that are {\em not} described by Breit-Wigner parametrization; 
instead they can be described by dispersion relations \cite{DISPERSION} in details (or other ways). At the qualitative level one should not be surprised. Probing the topologies of 
Dalitz plots with accuracy one might find the existence of ND.  Most of the data come along the frontiers. However, the centers are not empty; as we know 
direct CP asymmetries need interferences between (at least) two amplitudes, and the impacts of resonances are different for narrow vs. broad ones. We need more data, but also 
deeper thinking about underlying dynamics, whether those give us new lessons about non-perturbative QCD -- or also about ND. 

One looks at even more CKM suppressed three-body FS: 
\bea
\nonumber 
{\rm BR}(B^+ \to \pi^+\pi^-\pi^+) =(1.52 \pm 0.14 ) \cdot 10^{-5} \\
{\rm BR}(B^+ \to \pi^+K^-K^+) = (0.52 \pm 0.07 ) \cdot 10^{-5} 
\eea
It is not surprising that these rates are smaller than those in Eq.(\ref{KPIPI}). 
LHCb has measured these averaged \cp~asymmetries \cite{PRL112}:
\bea
A_{CP}(B^{\pm} \to \pi^{\pm} \pi^+\pi^-) &=& + 0.117 \pm 0.021_{\rm stat} 
\pm 0.009_{\rm syst}
[\pm 0.007_{\psi K^{\pm}}] 
\nonumber \\
A_{CP}(B^{\pm} \to \pi^{\pm} K^+K^-) &=& 
- 0.141 \pm 0.040_{\rm stat} \pm 0.018_{\rm syst} [\pm 0.007_{\psi K^{\pm}}] 
\label{AVERAGED}
\eea
As I have said above, re-scattering happen, although we have so far little quantitative control. 
Again, it is interesting that they come with opposite signs with only charged FS mesons like above in Eq.(\ref{SUPP1}), but they seems to be sizably larger than those.  
Maybe it is not `luck', but a pattern. On the other hand, it is not an obvious symmetry, but it depends 
on the situations. Penguin diagrams are suppressed, but they can produce large weak phases 
$b \to "W^-(t,c,u)" \to d$. It is a true challenge to understand its impact. 

LHCb has shown for `regional' \cp~asymmetries \cite{PRL112}:
\bea
\Delta A_{CP}(B^{\pm} \to \pi^{\pm} \pi^+\pi^-)|_{\rm regional} &=&  
+0.584 \pm 0.082_{\rm stat} \pm 0.027_{\rm syst}
[\pm 0.007_{\psi K^{\pm}}]  
\label{SUPP7} 
\nonumber \\
\Delta A_{CP}(B^{\pm} \to \pi^{\pm} K^+K^-)|_{\rm regional} &=&   
- 0.648 \pm 0.070_{\rm stat} \pm 0.013_{\rm syst}
[\pm 0.007_{\psi K^{\pm}}] \; .
\label{SUPP2}
\eea
Again it is not surprising that these asymmetries come with opposite signs. Maybe it might be somewhat 
surprising  that the impact on regional asymmetries are so large.  
We need more data, find other regional asymmetries and work on correlations with other FS. Importantly we need more thinking to understand what the data tell us about the underlying dynamics including 
non-perturbative QCD. Actually we have tested tools like dispersion relations \& chiral symmetry; 
however we have to apply them with more precision
It seems that the landscapes are more 
complex as said before and shows the impacts of broad resonances.  

\subsection{\ot~violation with \& without EPR correlations } 
\label{EPR}

\begin{figure}[h!]
\begin{center}
\includegraphics[scale=0.55]{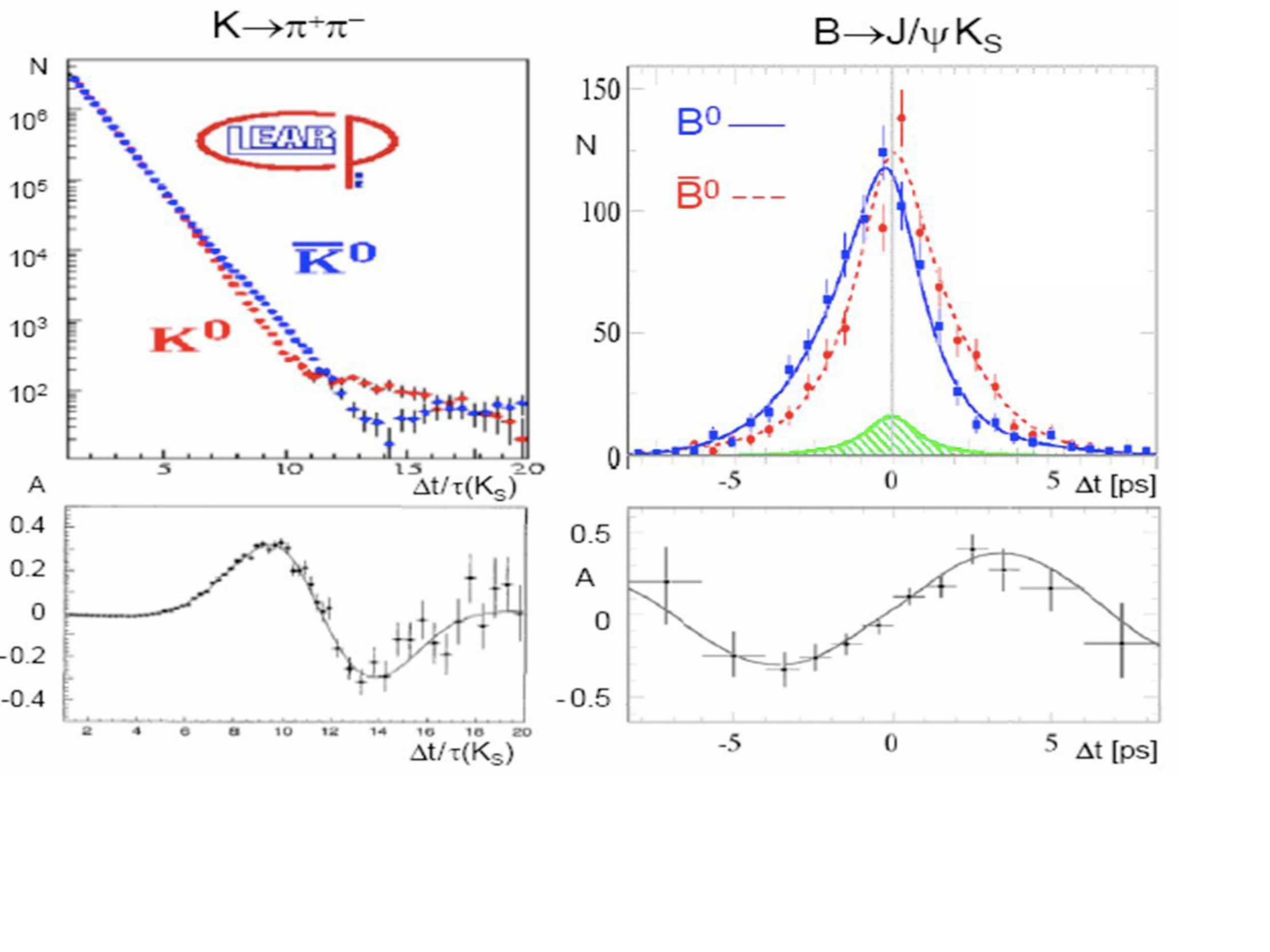}
\end{center}
\caption{The blue \& red data and the lines to describe them in QFT show in the right plot obvious difference of $B_d$ and $\bar B_d$ decays; furthermore they (within experimental uncertainties) are also $\Delta t$ odd. The landscape on the left 
side is complex: the difference is much smaller on $K^0$ vs. $\bar K^0$ 
on {\em average} than on the right side  [courtesy of K. Schubert].} 
\label{fig:COMPARE}
\end{figure}

Once one has established \CPV~directly, one has found 
\ot~violation indirectly with \cpt~invariance. However the situation is more subtle due to 
EPR correlation; actually it is a `blessing in disguise'.
People are not fans of history prefer the name of `entanglement' 
\footnote{`Entanglemant' seems to push out `EPR correlations' more and more recently in the 
literature; for me it 
is not only unfair, but worse by ignoring the history of quantum mechanics; furthermore it ignores to establish large \cp~asymmetries in $e^+e^- \to B_d \bar B_d$.}. For a special situation one has a pair of neutral $B$ mesons who are produce in 
single coherent quantum state with spin-one \& \oc~odd where their oscillations are highly correlated with 
each other as done at BaBar \& Belle experiments: $e^+e^- \to \Upsilon (4S) \to B_d \bar B_d$. 
This pair cannot transmogrify itself into a $B_dB_d$ or $\bar B_d \bar B_d$. To say it in 
different ways. Using the neutral mass eigenstates $B_1$ \& $B_2$\ one gets only  
$e^+e^- \to \Upsilon (4S) \to B_1B_2$, but not FS with $B_1B_1$ or $B_2B_2$.  The simplest and best measured asymmetry gives $\Upsilon (4S) \to (l^-X)_{B}(\psi K_S)_{B}$ vs. 
$\Upsilon (4S) \to (l^+\bar X)_{B}(\psi K_S)_{B}$ in the asymmetry of $e^+e^-$ collisions. One can measure 
the differences in the semi-leptonic \& non-leptonic decays. Those are depend on $\Delta t$, 
but also very consistent with sin$[\Delta M_{B_d}\Delta t]$ as expected. However there is more information, 
namely $\Delta t = 0$ within the experimental uncertainties. One has assumed \cpt~only for 
semi-leptonic decays, not non-leptonic one, as pointed out last century, shown on the Fig.\ref{fig:COMPARE}. 
The landscape of \cpt~violation has been probed with more details in Ref.\cite{BABARTV} -- but still assumes perfect EPR correlations.

\subsection{Weak decays of charm hadrons}
\label{CHARM}

No CP asymmetry has been found in charm mesons or baryons. On the other hand we have learnt that the 
landscape of charm transitions is complex in different directions. One is obvious, namely one can probe SCS \& DCS ones, where the SM gives small {\em weak} 
phases of ${\cal O}(0.001)$ in the first and basically zero on the latter. The second one is not so obvious: even when they depend on the same weak phase, they can be affected by 
(strong) re-scattering in different ways like two-, three- \& four-body FS. A well-known example: 
$\Gamma (D^0 \to K^+K^-)/\Gamma (D^0 \to \pi^+\pi^-) \sim 3$, while $\Gamma (D^0 \to K^+K^-\pi^+\pi^-)/\Gamma (D^0 \to 2\pi^+2\pi^-) \sim 1/3$. There are other examples with 
$D^0 \to \pi^+\pi^-$ \& $D^0 \to K^+K^- \pi^0$, $K_SK^-\pi^+$, $K_SK^+\pi^-$. 

\subsubsection{Indirect \CPV~in $D^0 - \bar D^0$ oscillation}
\label{DOSC}

$D^0 - \bar D^0$ oscillations have been established  by the data with 
$x_D\equiv\frac{\Delta M_D}{\bar \Gamma_D} =(0.39^{+0.17}_{-0.18} )\% $ and   
$y_D \equiv\frac{\Delta \Gamma_D }{2\bar \Gamma_D}  = (0.65^{+0.07}_{-0.09})\%$. 
The amplitude is described by SCS transitions, but also with the interference between Cabibbo favored 
\& DCS ones. The impact of ND can be seen mostly in $x_D$ due to local operators; the situation about $y_D$ is more complex \cite{BIANCO}. 

\subsubsection{Direct \CPV~in SCS decays of mesons}
\label{DIRCVD}

In the world of hadrons strong re-scattering connect $D \to \pi \pi \pi$ with 
$D \to \pi \bar K K$ and back. For very good reasons one describes three-body FS with 
amplitues with quasi-two body FS and their interferences; however scalar resonances often 
are described by broad ones where one cannot use Blatt-Wigner parametrization. Furthermore the Dalitz 
plots are not empty; therefore interferences happens at many locations. The SM is expected to produce 
averaged values for SCS decays~ ${\cal O}(0.001)$ and larger values for regional asymmetries. The questions are: how much, where and about the impact of \cpt~invariance   
on subclasses with only charged hadrons or not.   One has to probe averaged \& regional ones in $D_s^{\pm} \to K^{\pm}K^+K^-, K^{\pm}\pi^+\pi^-$ and to think about correlations with $D^{\pm}$ decays. 

Chiral symmetry is a very good tool for $3\pi$ FS; however the power of that is decreased for 
FS with $K\pi\pi$, $K\bar K \pi$ and $3K$. Again -- how much?

\subsubsection{Basically zero \cp~asymmetries in DCS decays}
\label{DCS}

The refined parameterization of the CKM matrix \cite{AHN}  gives basically zero direct \CPV~in DCS in 
$D^{\pm} \to K^{\pm}\pi^+\pi^-/K^{\pm}K^+K^-$ and in exotic decay 
$D_s^{\pm} \to K^{\pm}K^{\pm}\pi^{\mp}$. The first step is to establish averaged \CPV~in $D_{(s)}$ decays, 
then the second one is to probe regional ones. Again it needs some judgment to define regional 
asymmetries with finite data.
While the rates are very small, there is no `background' 
from the SM. Furthermore the `exotic' $D_s$ decays should be more standing out 
due to $\Delta S=2$ in the FS; at least they give us unusual lesson about QCD. 

I add a comment about CP asymmetries in the decays of charm baryons like $\Lambda^+_c$. 
One can compare favored decays $\Lambda_c^+ \to p K^-\pi^+$ with DCS $\Lambda_c^+ \to p K^+\pi^-$: 
\bea 
{\rm BR}(\Lambda_c^+ \to p K^-\pi^+) &=& (6.84 ^{+0.32}_{-0.40}) \cdot 10^{-2}    \\
{\rm BR}(\Lambda_c^+ \to p K^+\pi^-) &<&  3.1 \cdot 10^{-4}
\eea 
The favored decay has been established. For the future it should be transformed from "accuracy" to "precision". 
The latter one has not been founded yet; 
the first step is to establish with some accuracy. The second step would be to probe CPV in $\Lambda_c^+ \to p K^+\pi^-$ as calibrated by 
$\Lambda_c^+ \to p K^-\pi^+$, where there is hardly a chance to find CPV in general. 
It would be an excellent achievement to find CPV due to two points: to establish CPV in baryon decays for the first time -- and also to find impact of ND at the 
same time without background from the SM. 
Then the third step would be to probe regional CP asymmetries. 

\subsection{Weak decays of beauty baryons}
\label{BEAUTYBARYON}

I have said it  before that it is important to probe {\bf CP} asymmetries in the decays of beauty baryons. For example, I gave two talks to the LHCb collaboration 
at CERN in June 2016 about these items: compare the rates of $\Lambda_b^0 \to p \pi^-$ vs. $\bar \Lambda_b^0 \to \bar p \pi^+$  \& 
$\Lambda_b^0 \to p K^-$ vs. $\bar \Lambda_b^0 \to \bar p K^+$ and the Dalitz plots for $\Lambda_b^0 \to \Lambda^0 \pi^+\pi^-, \Lambda^0 K^+K^-, \Lambda K^+\pi^-$ vs. 
$\bar \Lambda_b^0 \to \bar \Lambda^0 \pi^+\pi^-, \bar \Lambda^0 K^+K^-, \bar \Lambda K^-\pi^+$. And 
to measure the angles between two planes for $\Lambda_b^0 \to p \pi^-\pi^+\pi^-$ vs. $\bar \Lambda_b^0 \to \bar p \pi^+\pi^-\pi^+$ \& 
$\Lambda_b^0 \to p \pi^-K^+K^-$ vs. $\bar \Lambda_b^0 \to \bar p \pi^+K^-K^+$ with different definitions of their planes. It helps 
to deal with production asymmetries 
in $pp$ collisions and go after the impact of ND. I am not surprised that LHCb will very soon send out a paper about the moment of $\Lambda_b^0 \to p \pi^-\pi^+\pi^-$ with a 
3.3 $\sigma$ uncertainties away from zero \cite{LHCbtalk}. It would be wonderful achievement on the experimental side, when it is established. Of course, it would be very good also going 
after the situation with $\Lambda_b^0 \to p \pi^-K^+K^-$ to use {\bf CPT} invariance. On the negative side: the value of around 20 \% for direct {\bf CP} violation is too large. That is life on several levels?

\section{Intermezzo --  electric dipole moment (EDM)}
\label{EDM}

It is a rich landscape that shows the connection of HEP, Hadrodynamics, nuclear, atomic \& molecule physics. 
\begin{itemize}
\item
We have found large \CPV~flavor dynamics, but it has nothing to do with the truly huge observed 
asymmetry in known matter vs. anti-matter of baryons. 
\item 
EDMs have been probed in many different situations \cite{EDMSCERN}, but none has been found yet despite hard work both on the experimental \& theoretical side. 
Still I see no good reason to give up: future work might tell us the direction for the dynamics to produce that asymmetry.

\item 
QCD faces the challenge to solve the problem with basically zero 
contributions of the operator 
$G\cdot \tilde{G} \equiv i \epsilon_{\mu \nu \alpha \beta} G_{\mu \nu} G^{\alpha \beta}$ 
and the gateway for `traditional' and `novel' axions.

\item 
There might be a connection of known vs. dark matter.   
\item 
A very general statement: to understand fundamental dynamics it needs a lot of time, new tools -- 
and thinking \&  new ideas.  

\end{itemize}

Direct test of \ot~invariance comes for single particle static transitions. The energy shift of a system due 
to external small electric field can be described in powers of $\vec{E}$: 
\beq
\Delta {\cal E} = \vec{d}\cdot  \vec{E} + {\cal O}(|\vec{E}|^2) = d\, \vec{j}\cdot \vec{E} + {\cal O}(|\vec{E}|^2)
\eeq
The linear vector $d_i$ is called the EDM. A non-zero value of $d_i$ 
show the violation both discrete \op~and \ot~symmetries. The crucial point is not `elementary', but 
`non-degeneracy' of the impact of the dynamics. It is a well-known example to compare the 
neutron $d_N$ with water molecules or `dumb-bells' based on classical forces \cite{CPBOOK}. 

In quantum field theory EDMs are described by an operator in the Lagrangian 
\beq
{\cal L}_{\rm EDM} = - \frac{i}{2}\, d \, \bar \psi \sigma_{\mu \nu}\gamma_5\psi F^{\mu \nu}
\eeq
with dimension five, while the Lagrangian has dimension four; therefore its dimensional coefficient $d$ can be calculated as a finite quantity in general. 

In the SM one gets EDM values for neutrons, deuterons, molecules and also for $e$, $\mu$ \& $\tau$ ones that are clearly beyond what one can reach. Therefore it 
is a rich landscape for the existence of ND and its features, if you are patient enough to make 
the efforts that are needed with thinking about \& working on ideas. 

The situation is more subtle: QCD can produce \CPV~in flavor independent transitions, namely EDMs in hadrons. 
It was realized long time ago that 
QCD with vector bosons have a problem used `$U(1)_A$ problem' \cite{CPBOOK}. Let us look at QCD with only one 
family with $u$ \& $d$ quarks. With massless quarks -- which is very close to 
$m_u$, $m_d$ $\ll \bar \Lambda$ -- one might think that QCD possesses a {\em global} 
$U(2)_L \times U(2)_R$ symmetry. Indeed the vectorial component $U(2)_{L+R}$ is conserved even after 
QM corrections and axial $SU(2)_{L-R}$ also in subtle ways to give masses to $W_{\mu}^{\pm}$ and 
$Z^0$. However about $U(1)_{L-R}$? Axial currents are conserved in the classical symmetry due to 
chiral invariance for massless quarks; however they are not conserved called `quantum anomaly' 
(or `triangle anomaly') due to one-loop corrections with internal quarks:
\beq
\partial _{\mu}J_{\mu}^5 = \frac{g_S^2}{32\pi^2} G\cdot \tilde{G}
\eeq
The resolution of the $U(1)_{L-R}$ due to complex structure of the QCD `vacuum' comes with a price, 
namely the `Strong \cp~Problem'. The $U(1)_{L-R}$ \& Strong \cp~is actually intertwined, 
when one includes the weak dynamics \footnote{Often our community is sloppy with the names 
understanding the connections; other examples below: KSVZ or DFSZ axion.}: 
\beq
{\cal L}_{\rm eff} = {\cal L}_{\rm QCD} + \frac{\bar \theta g_S^2}{32\pi^2} G\cdot \tilde{G}
\eeq
with the observable $\bar \theta = \theta - {\rm arg \, det} {\cal M}$. It describes the mixing matrix of 
U=(t,c,u) and D=(b,s,d) quarks. 
Photon can couple neutrons with internal virtual protons \& pions. One of the two effective pion nucleon operators couple by ordinary QCD, 
while to other one are due $G\cdot \tilde{G}$. A guess tell us: 
\beq
d_N \sim {\cal O}(10^{-16}\bar \theta) {\rm e\; cm}
\eeq 
The limit from the data gives: 
\beq
\bar \theta < 10^{-10}
\eeq
While it is possible or worse `accidentally', but very `un-natural'.

\subsection{Traditional \& novel axion scenarios}
\label{AXION}

Most members of our community agree that one needs organizing principle to produce the 
required cancellations. The best known tool is some kind of symmetry. A real intriguing ansatz is 
to assume that a physical quantity usually used as a constant is re-interpret as a dynamics degree 
of freedom. In this case it was suggested by Peccei \& Quinn \cite{PQ}, namely to add the SM a 
global $U(1)_{\rm PQ}$ as a Nambu-Goldstone boson which is axial with following properties: 
\begin{itemize}
\item 
it is a classical symmetry; 
\item 
it is subject to an axial anomaly;   

\item
it is broken spontaneously as well and 
\item
possesses a huge vacuum expectation value (VEV) $v_{\rm PQ} >> v_{\rm EW}$.

\end{itemize}
Previously we thought there are two classes, namely (A) `visible' axion with 
$m_a \sim {\cal O}(1\, {\rm MeV})$, and (B) `invisible' axion with $m_a \ll 1\, {\rm MeV}$. It seems 
there is no chance that class (A) axion can exist. `Invisible' axion might be found using coupling of axion with two photons. The name `invisible' is obvious, 
namely due to the tiny axion mass the lifetime is larger than the age of our universe, and couplings to other fields are so minute that they would not betray their presence under `ordinary' circumstances. 
The best tool might be by conversion an axion into a photon in a strong 
magnetic field {\bf B} \cite{SIK}: 
\beq 
{\rm axion}
\; \;   \stackrel{\bf B}{\Longrightarrow}  \; \;  {\rm photon} 
\eeq 
It is still probed in our present world (including solar system due to astrophysical indirect information). 

Later `old' cosmology enters the `scene': it gave lower bound on the mass: 
\beq
m_a > 10^{-6} \; {\rm eV}
\eeq
Then connections of dark matter suggest stronger bounds: 
\beq
m_a > 2 \cdot 10^{-5} \; {\rm eV}
\eeq
Does it mean that the `dawn' of axions goes to the `dusk'? 
Maybe the landscape of axion dynamics is even more subtle; 
namely actually PQ symmetry can be broken not only in QCD anomaly, but also 
in the UV region in many ways (and ideas) due to connection with gravity, gravitational waves,  
string theoretical realization of the QCD axion etc. etc. Axions produced in the very early universe, 
can be part of the Dark Matter (and maybe also in Dark Energy) in the present universe and can be tested {\em experimentally} and {\em directly}. 
For example, it was described with more observables in the Refs.\cite{KIWOON},\cite{RAJENDRAN},\cite{JKIM} 
about the PLANCK \& BICEP2 data. 

I am not convinced (yet) by some comments; however even if those projects will not be  
realized, they show 
the active situation in fundamental physics, which is wonderful in my view: a true `Renaissance' 
from an excellent idea about the impact of symmetries.

\section{Probing \cp~asymmetries in leptonic transitions}
\label{PMNS}

In the SM the landscape of \CPV~in hadrons and leptons are quite different. There the charged 
leptons and neutrinos are elementary with no original \CPV. This century  
 data showed us that neutrinos are not massless due to oscillations. Some of us think that \CPV~in neutrino 
oscillations can show the road to understand the huge difference between 
matter vs. antimatter. It also shows we need a very long time efforts to make progress there. 

The SM landscape of leptonic dynamics about \CPV~ is not very complex with massless neutrinos and 
$e$, $\mu$ and $\tau$ transitions. One can see it as not very interesting -- or opposite, since 
there is hardly SM background on the theoretical side.

\subsection{$\tau$ Cabibbo suppressed decays}
\label{TAUDECAY}

Present data about \CPV~in SCS $\tau$ decays 
$\tau ^- \to \nu K_S (\pi ...)^-$   show one can compare SM prediction due to well-known 
$K^0 - \bar K^0$ oscillation with a difference of 2.9 sigma: 
\bea
A_{\rm CP}(\tau ^+ \to \bar \nu K_S\pi^+)|_{\rm SM} &=& +(0.36 \pm 0.01) \% \; \; \; \cite{TAUBS} 
\label{TAU1}
\\
A_{\rm CP}(\tau ^+ \to \bar \nu K_S\pi^+ [+\pi^{0\; \prime} {\rm s}])|_{\rm BaBar2012} &=& -(0.36 \pm 0.23 \pm 0.11) \%  \; ;
\; \; \cite{BABARTAU}
\label{TAU2}
\eea
one can note the sign. One can probe \CPV~decays like $\tau^- \to \nu K^-\pi^0$, 
$\nu K^-\pi^+\pi^-$ etc. and think about correlations due to \cpt. We have to probe 
\CPV~in several FS like $\tau ^- \to \nu K^-\pi^0$, $\nu K^-\pi^+\pi^-$, 
$\nu K_S\pi^- \eta$. 

Now available data probe only integrated CP asymmetries. It is important to probe 
regional CP asymmetries in $\tau ^- \to \nu [S=-1]$ FS; we have to wait for Belle II 
(and Super-Tau-Charm Factory if \& when it exists). Furthermore one has to compare 
regional data from $\tau ^- \to \nu [S=0]$ FS like $\tau ^- \to \nu \pi^-\pi^0$, 
$\nu \pi^-\eta$, $\nu \pi^-\pi^+\pi^-$, $\nu \pi^-\pi^0\pi^0$ etc. with accuracy. 
It is a test of 
experimental uncertainties; it would be a miracle to show \CPV~there.

It is important (as pointed out two years ago) to measure the correlations with 
$D^+ \to K^+\pi^+\pi^-/K^+K^+K^-$ etc. \cite{TAUD+}. 
Furthermore we have to look for regional asymmetries and spin correlations in the pairs 
of $\tau^+\tau^-$, in particular with polarized $e^+e^-$ beams if we can use them.

\subsection{\CPV~in neutrino oscillations}
\label{NEUTRINO}

PMNS matrix very different than CKM matrix already in qualitative ways. In the world of quarks and also 
charged leptons masses they follow the catholic hierarchy.  The situation is quite different about 
neutrino masses and angles. Furthermore neutrinos might have be partly Majoran. In general 
the three angles of the PMNS matrix \cite{PMNS} differ sizably from zero. 

It is a very long time project to probe \CPV~in neutrino oscillations, which are affected by the 
environment of very mostly baryons rather than anti-baryons.

\section{Summary and about the Future}

Now we have entered the era where `accuracy' has been changed into `precision' with better tools including much better understanding 
of strong forces -- and the possible connection with dark matter.

Up to now \CPV~basically have been measured in two-body FS in kaons and 
$B$ mesons. It is crucial to probe many-body FS in kaons, $D_{u,d,s}$ and $B_{u,d,s}$ and 
in baryons in general. Furthermore we have to use 
\cpt~as a tool to connect informations from different FS and regional \CPV. 
EDMs in nuclei \& molecules show us a new road for ND even, when it is not connected with the asymmetry in matter vs. anti-matter. 

No \CPV~or \TV~ has been found in leptonic dynamics with small limits so far. 
However we have to continue with precision, not only to understand those, but have a chance to find the source of the huge asymmetry 
in baryons vs. anti-baryons. Finally we know that the SM is not enough in our universe due to dark matter existence \& 
neutrino oscillations. 
Therefore we have to probe \CPV~in neutrino oscillations, although it is a true challenge 
where we need long time projects based on HEP, Hadrodynamics \& Nuclear Physics and combine their informations.

\vspace{0.5cm}

{\bf Acknowledgments:} 
This article has been published by BENTHAM OPEN, {\em Open Physics Journal}, 2016, {\em 3}, 55 - 76. . 
There is only one addition: Sect.3.11 with the LHCb data for {\bf T}-odd moment shown at ICHEP2016 in Chicago.

This work was supported by the NSF under the grant numbers PHY-1215979 and 
PHY-1520966. 
\vspace{4mm}


\end{document}